\documentstyle[prl,aps]{revtex}
\begin{document}
\draft
\title{$\Phi-$measure of  azimuthal fluctuations
%\footnote{nucl-th/9907099}
}
\author{Stanis\l aw Mr\' owczy\' nski\footnote{Electronic address:
{\tt mrow@fuw.edu.pl}}}

\address{So\l tan Institute for Nuclear Studies, \\
ul. Ho\.za 69, PL - 00-681 Warsaw, Poland \\
and Institute of Physics, Pedagogical University, \\
ul. Konopnickiej 15, PL - 25-406 Kielce, Poland}

\date{24-th July 1999, revised 2-nd September 1999}

\maketitle

\begin{abstract}

The event-by-event azimuthal fluctuations in high-energy heavy-ion
collisions are analyzed by means of the so-called $\Phi-$measure.
The fluctuations due to the collective transverse flow and those 
caused by the quantum statistics and resonance decays in the hadron 
gas are discussed in detail.

\end{abstract}

\vspace{0.5cm}
 

\vspace{0.5cm}

%\newpage

Various phenomena can lead to the nontrivial azimuthal  fluctuations 
in (ultra-)relativistic heavy-ion collisions. One mentions here jets and 
minijets being the result of  (semi-)hard parton-parton scattering 
\cite{Wan97} and the collective transverse flow due to  
the anisotropic pressure gradient \cite{Ame91,Oll92}. The latter 
phenomenon, which is well known at the intermediate collision energies 
(for a review see e.g. \cite{Rei97}), has been recently observed in the 
nucleus-nucleus collisions at CERN SPS \cite{Agg97,App98}. As argued 
in the series of our papers \cite{Mro93}, the color plasma instabilities 
also generate a transverse collective flow in heavy-ion collisions at 
RHIC and LHC. Since the final state azimuthal fluctuations appear 
as remnants of  the inhomogeneous or anisotropic early stage of  the 
collision, when the minijets are copiously produced \cite{Gyu97} or the 
color instabilities occur \cite{Mro93}, the effects of interests are expected 
to be rather small \cite{Wan98}. Thus, it is a real challenge to extract 
them from the statistical noise. 

A method, which has been successfully applied to study the collective flow
in heavy-ion collisions at SPS energies \cite{App98}, is based on the Fourier 
analysis of the azimuthal distribution \cite{Vol96,Pos98,Oll97}. Specifically, 
one expands the distribution of the particle azimuthal angle measured with 
respect to the reaction plane, which is reconstructed on the event-by-event 
basis, into the Fourier series. Then, the first harmonic tells us about the 
so-called directed flow, the second one about the elliptic flow, etc. Although 
the method is very powerful there are subtleties in applying it to the real data. 
In particular, the finite statistics and non-flow interparticle correlations - the 
jets are expected to be an important source of such correlations at RHIC - distort 
the reconstructed event plane \cite{Pos98}. Due to the finite resolution of the 
reconstruction procedure, it is difficult to measure the Fourier coefficients higher 
than the second one \cite{Oll97}. 

The Fourier expansion method has been designed to study the collective 
flow which is correlated with the reaction plane. However, the jets or color 
instabilities, which have been mentioned above, generate the transverse 
collective motion being, at least approximately, independent of the 
reaction plane orientation. Then, the Fourier expansion method, as developed
in \cite{Vol96,Pos98,Oll97}, is not sensitive to such phenomena and one needs
another tools for the data analysis, for example the method which focuses 
on the two-particle large-angle azimuthal correlations \cite{Wan91}.

In this letter we propose to study the azimuthal fluctuations by means of  
the previously introduced fluctuation measure $\Phi$ \cite{Gaz92}. Although 
the measure has been invented for other purposes, see below, it might be also 
useful for the case of azimuthal fluctuations. The $\Phi-$measure analysis 
does not demand the reaction plane reconstruction and can be easily applied 
to the data. $\Phi$ is also sensitive to the different sources of  correlations. It appears 
difficult to disentangle various contributions but the integrated information 
provided by $\Phi$ can be combined with that offered by other methods, in 
particular the Fourier expansion. For example, we show in this note that 
the $\Phi-$measure of $\phi-$fluctuations deviates from zero when the 
transverse collective flow is present in the system. Since all harmonics contribute 
to $\Phi$ one can check whether the first two Fourier coefficients, which are
measured, saturate the observed value of $\Phi$. The analysis of $\Phi-$measure 
of $\phi-$fluctuations can be also combined with the $\Phi-$measure studies 
of other kinematical variables. $\Phi$ of $p_T-$fluctuations has been already shown 
to be very sensitive to the presence of  the jets \cite{Liu99}. Thus, the simultaneous 
measurement of $\phi-$ and $p_T-$fluctuations can be helpful in disentangling 
various contributions. One should remember however that except the dynamical 
correlations caused by the hydrodynamic flow, jets or color instabilities, there 
are also correlations due to the quantum statistics of particles. These correlations 
are always present. Therefore, we compute $\Phi$ in the ideal pion gas to estimate 
how the Bose statistics influences the azimuthal fluctuations. We also consider 
here the effect of hadron resonances. 

Let us introduce the correlation (or fluctuation) measure $\Phi$. 
We define the variable $z\buildrel \rm def \over = x - \overline{x}$,
where $x$ is a single particle characteristics such as the particle transverse 
momentum or the azimuthal angle. The overline denotes averaging 
over a single particle inclusive distribution. In our further considerations, 
$x$ is identified with the particle azimuthal angle. The event variable $Z$, 
which is a multiparticle analog of $z$, is defined as 
$Z \buildrel \rm def \over = \sum_{i=1}^{N}(x_i - \overline{x})$, where the 
summation runs over particles from a given event. By construction, 
$\langle Z \rangle = 0$, where $\langle ... \rangle$ represents averaging 
over events (collisions). The measure $\Phi$ is finally defined in the 
following way
\begin{equation}\label{phi}
\Phi \buildrel \rm def \over = 
\sqrt{\langle Z^2 \rangle \over \langle N \rangle} -
\sqrt{\overline{z^2}} \;.
\end{equation}
As shown in \cite{Gaz92}, $\Phi-$measure possesses properties which are
very useful in the data analysis. It equals zero when the final sate particles 
are independent from each other i.e. there is no inter-particle correlation in the 
system. The second property is less trivial: the value of $\Phi$ is exactly the 
same for the nucleon--nucleon (N--N) and nucleus--nucleus (A--A) collisions, 
if the A--A case is a superposition of N--N interactions with no secondary 
scatterings. In other words, the strength of the correlation is not influenced 
by the number of uncorrelated particle sources. The measure $\Phi$ has 
been successfully applied to the NA49 experimental data and it has been 
found \cite{Rol98,App99} that the dynamical transverse momentum 
correlations, which are present in N--N collisions, are significantly 
reduced in the central Pb--Pb reactions. The correlations observed in
these collisions are fully explained by the effect of Bose statistics 
of pions \cite{Mro98,Mro99b}.

Let us compute the $\Phi-$measure of the azimthal fluctuations caused
be the transverse collective flow. The inclusive $\phi-$distrubtion is,
of course, flat i.e.
\begin{equation}\label{incl}
P_{inc}(\phi) = {1 \over 2\pi} \; \Theta(\phi) \, \Theta(2\pi - \phi) \;,
\end{equation}
which gives $\overline \phi = \pi$ and 
$\overline{\phi^2} = {4 \over 3}\pi^2$, and consequently,
$\overline{z^2} = {1 \over 3}\pi^2$. Following \cite{Vol96}, the azimuthal 
distribution of a single event is chosen in the form of the Fourier series i.e.
\begin{equation}\label{event1}
P_{ev}(\phi) = {1 \over 2\pi} \;
\Big[1 + \sum_{n=1}^{\infty} 
\big( X_n {\rm cos}(n\phi ) + Y_n {\rm sin}(n\phi ) \big) \Big] 
\Theta(\phi) \, \Theta(2\pi - \phi) \;,
\end{equation}
where the parameters $X_n$ and $Y_n$ change from event to event. The 
distribution (\ref{event1}) is usually rewritten as
\begin{equation}\label{event2}
P_{ev}(\phi) = {1 \over 2\pi} \;
\Big[1 + 2\sum_{n=1}^{\infty} v_n {\rm cos}\big(n(\phi- \psi_n)\big) \Big] 
\Theta(\phi) \, \Theta(2\pi - \phi) \;,
\end{equation}
where 
$$
X_n = 2v_n{\rm cos}(n\psi_n)  \;,\;\;\;\;\;\;\;\;\;\;
Y_n = 2v_n{\rm sin}(n\psi_n ) \;.
$$
The first term of the Fourier series from eq. (\ref{event1}) or (\ref{event2}) 
corresponds to the so-called transverse directed flow. The amplitude $v_1$ 
controls strength of the flow while $\psi_1$ determines the reaction plane. 
The meaning of  $v_n$ and $\psi_n$ for $n > 1$ is analogous.

The angles $\psi_n$ vary from event to event and their distributions are flat. 
Therefore, the event distribution (\ref{event2}) averaged over events provides, 
as it should, the inclusive distribution (\ref{incl}). Further, we consider two 
extreme cases. In the first one, which seems to be appropriate for the 
hydrodynamic flow analysis, the $\psi_n$ angles are maximally correlated to 
each other and uniquely determined by the reaction plane angle $\psi_r$ i.e. 
$\psi_r = n\psi_n + \alpha_n$. Then, the averaging over events corresponds 
to the integration over the angle $\psi_r$. In the second case, the angles 
$\psi_n$ are independent from each other and one integrates over all 
$\psi_n$ to average over events.

Since $Z = \sum_{i=1}^{N}(\phi_i - \overline \phi)$ one finds in the first 
case that
\begin{eqnarray}\label{Z2}
\langle Z^2 \rangle = {1\over 2\pi} \int_0^{2\pi} d\psi_r \sum_N {\cal P}_N
\int_0^{2\pi} d\phi_1 .....\int_0^{2\pi} d\phi_N\; 
P_{ev}(\phi_1)\; .... \; P_{ev}(\phi_N) \;
(\phi_1 + .....+ \phi_N - N\overline \phi )^2 \;,
\end{eqnarray}
where ${\cal P}_N$ is the multiplicity distribution. The formula analogous to 
(\ref{Z2}), which corresponds to the second case,  includes the averaging over 
all angles $\psi_n$. It is understood that there is one more averaging in 
eq.~(\ref{Z2}) which is not explicitly shown. Namely, the averaging over
the amplitudes $v_n$ which also change from event to event. To simplify
the notation we also neglect here a correlation between the event 
multiplicity and the flow strength. After elementary calculation, one finds 
from eq.~(\ref{Z2}) that 
$$
\langle Z^2 \rangle = {\pi^2 \over 3} \langle N \rangle 
+ \big(\langle N^2 \rangle  - \langle N \rangle \big) \, S \;,
$$
with
$$
\langle N^k \rangle \buildrel \rm def \over = \sum_N N^k {\cal P}_N \;,
$$
and 
\begin{displaymath}
S = 2 \times \left\{ 
\begin{array}{ccl}
\langle \: \sum_{n=1}^{\infty} \big({v_n \over n} \big)^2 \rangle
&{\rm for}& {\rm 1\!-\!st \; case,} \\[2mm]
\langle \big( \sum_{n=1}^{\infty} {v_n \over n} \big)^2 \rangle
&{\rm for}& {\rm 2\!-\!nd \; case.} 
\end{array}
\right.
\end{displaymath}
Finally, we get 
$$
\Phi = \sqrt{ {\pi^2 \over 3} +  \bigg({\langle N^2 \rangle  
- \langle N \rangle \over \langle N \rangle }\bigg)\, S }
- {\pi \over \sqrt{3}}\;.
$$
As expected, $\Phi = 0$ for $S=0$. When $S \rightarrow 0$,
$\langle N^2 \rangle \cong \langle N \rangle^2$ and  
$\langle N \rangle \gg 1$, we get an approximate expression
\begin{equation}\label{phi-col}
\Phi \cong {3 \over 2 \pi^2} \, \langle N \rangle \, S \;.
\end{equation}

Let us consider how large is the expected effect. The amplitudes $v_1$ and 
$v_2$ observed in Pb--Pb collision at 158 GeV per nucleon do not exceed 
for pions the value of 0.03 \cite{App98}. We take $v_1 = v_2 = 0.03$ and 
$v_n=0$ for $n > 2$. We also neglect here the variation of $v_1$ and $v_2$.
Then, one finds from Eq.~(\ref{phi-col}) that for $\langle N \rangle = 170$ 
\cite{App98} $\Phi$ equals 0.058 in the first case and 0.105 in the second one.

As already mentioned, the transverse flow is far not the only source
of the azimuthal fluctuations. In particular, the quantum correlations 
contribute here. We compute $\Phi$ in the ideal quantum gas to 
estimate the effect of quantum statistics. Modifying our previous 
calculations \cite{Mro98,Mro99a,Mro99b}, one immediately finds
\begin{equation}\label{phi-ther}
{\langle Z^2 \rangle \over \langle N \rangle }= 
{1 \over \rho}\int{d^3p \over (2\pi )^3}
\,(\phi - \overline{\phi})^2 \; {\lambda^{-1}e^{\beta E}
\over (\lambda^{-1}e^{\beta E} \pm 1)^2} \;,
\end{equation}
where 
\begin{equation}\label{rho}
\rho = \int{d^3p \over (2\pi )^3} \;
{1 \over \lambda^{-1}e^{\beta E} \pm 1} \;;
\end{equation}
$\beta \equiv T^{-1}$ is the inverse temperature; 
$\lambda \equiv e^{\beta \mu}$ denotes the fugacity and $\mu$ the chemical 
potential; $E$ is the particle energy equal to $\sqrt{m^2 + {\bf p}^2}$ with 
$m$ being the particle mass and ${\bf p}$ its momentum; the upper sign is 
for fermions while the lower one for bosons. 

Since the inclusive azimuthal distribution is again given by (\ref{incl}), 
we get 
\begin{equation}\label{phi-phi}
\Phi = {\pi \over \sqrt{3}}
\bigg( \sqrt{{\widetilde \rho \over \rho}} - 1 \bigg)\;,
\end{equation}
where
\begin{equation}\label{rho-til}
\widetilde \rho = \int{d^3p \over (2\pi )^3} \;
{\lambda^{-1}e^{\beta E}
\over (\lambda^{-1}e^{\beta E} \pm 1)^2} \;.
\end{equation}

As seen, $\Phi$ is an intensive thermodynamic quantity, i.e. it is 
independent of the system volume. It is also independent of the
number of the particle internal degrees of freedom. One easily 
observes that $\Phi < 0$ for fermions, $\Phi > 0$ for bosons 
and $\Phi = 0$ in the classical limit ($\lambda^{-1} \gg 1$).

When the particles are massless and their chemical potential vanishes
($\lambda = 1$), $\Phi$ can be calculated analytically and the result 
reads
$$
\Phi = {\pi \over \sqrt{3}} \bigg(  
\sqrt{{\pi^2 \over 6 \zeta(3)}{2/3 \choose 1}} - 1 \bigg)
\cong {-0.082 \choose \;\;\;0.309} \;,
$$
where $\zeta(x)$ is the Riemann zeta function  with $\zeta(3) \cong 1.202$.

In Fig. 1 we present with dashed lines the $\Phi-$measure of 
$\phi-$fluctuations in the ideal pion gas as a function of temperature. 
The pions are, of course, massive ($m_{\pi}=140$ MeV). The calculations 
are performed for several values of the pion chemical potential. The 
chemical equilibrium corresponds to $\mu = 0$. As seen, $\Phi$ grows 
with the temperature. In the classical limit, when $\mu \rightarrow 0$, 
the $\Phi-$measure vanishes.

It is a far going idealization to treat a fireball at freeze-out 
as an ideal gas of pions. A substantial fraction of the final state 
pions come from the hadron resonances. These pions do not `feel' the 
Bose-Einstein statistics at freeze-out and consequently the value
of $\Phi$ should be significantly reduced. We estimate the role of 
resonances as in our earlier papers \cite{Mro99a,Mro99b}. The spectrum of 
pions, which originate from the resonance decays, is not dramatically 
different than that given by the equilibrium distribution \cite{Sol91}. 
Therefore, we treat the fireball at freeze-out as a mixture of `quantum' 
pions - those called `direct' - and the `classical' ones which come from 
the resonance decays. The $\Phi-$measure is again given by Eq.~(\ref{phi-phi}) 
but the formulas (\ref{rho},\ref{rho-til}) are modified as
$$
\rho = \int{d^3p \over (2\pi )^3} \;
\bigg[{1\over \lambda^{-1}e^{\beta E} - 1} 
+ \lambda_r e^{-\beta E} \bigg] \;,
$$
$$
\widetilde \rho = \int{d^3p \over (2\pi )^3} \;
\bigg[{\lambda^{-1}e^{\beta E}
\over (\lambda^{-1}e^{\beta E} - 1)^2}
+ \lambda_r e^{-\beta E} \bigg] \;.
$$
The parameter $\lambda_r$ is chosen is such a way that the number of 
`classical' pions equals the number of pions from the resonance
decays. Thus, $\lambda_r$ is temperature dependent. In the actual 
calculations we have taken into account the lightest resonances:  
$\rho(770)$ and $\omega(782)$ which give the dominant contribution.
The life time of $\rho$, which is 1.3 fm/$c$, is not much longer
than the time of the fireball decoupling and some pions from 
the $\rho$ decays can still `feel' the effect of Bose statitistics. 
Therefore, the contribution of $\rho$ to the `classical' pions is 
presumably overestimated in our calculations. Since we neglect the 
heavier resonances  and weakly decaying particles, which also contribute
to the final state pions,  the two effects partially compensate each
other.  In any case, our calculations show that the resonances do
not change the value of $\Phi$ dramatically in the domain of 
temperatures of interest. 

In Figs. 1 the solid lines represent $\Phi-$measure which includes 
the resonances. The chemical potentials of $\rho$ and $\omega$ are 
assumed to be equal to that of pions. As seen, the role of the 
resonances is negligible at the temperatures below 100 MeV but above 
this temperature the resonances reduce the fluctuations noticeably.
The freeze-out temperature in Pb--Pb collisions at 158 GeV per nucleon, 
which is obtained by means of the simultaneous analysis of the single 
particle spectra and the two-particle correlations, is about 120 MeV  
\cite{App98a}. For $T=120$ MeV and $\mu=0$ the $\Phi-$measure 
equals 0.078, when the resonances are neglected, and is reduced to 0.066  
when the resonances are taken into account. One observes that the effects 
of  the quantum statistics and transverse flow are of comparable size. 

The $\Phi-$measure given by Eq.~(\ref{phi}) corresponds to the 
second moment of the fluctuating quantity. It has been suggested 
\cite{Bel99} to use the higher moments in an analogous way. However, 
we have shown \cite{Mro99b} that only the third moment measure preserves 
the advantageous properties of $\Phi$ while the higher moment measures 
do not. We have also argued \cite{Mro99b} that the simultaneous usage 
of $\Phi_2$ and $\Phi_3$ may help in identifying the origin of correlations 
observed in the final state of heavy-ion collisions. Unfortunately, the third 
moment measure is useless in the studies of $\phi-$fluctuations. One easily 
shows that due to the symmetry $\overline{z^3} = 0$ and $\langle Z^3 \rangle = 0$ 
when the variable $x$ is identified with the azimuthal angle.

We conclude our considerations as follows. The $\Phi-$measure, which 
can be easily applied to the experimental data, seems to be a useful tool to 
analyze the azimuthal fluctuations in heavy-ion collisions. It is sensitive to 
the different nontrivial fluctuations, in particular those caused by the transverse 
flow and quantum statistics which have been quantitatively discussed here. 
The $\Phi-$measure analysis combined with other techniques, such as 
the Fourier expansion method \cite{Vol96,Pos98,Oll97}, will help to
study various sources of the fluctuations. 

\vspace{1cm}

I am very grateful to Marek Ga\'zdzicki, Art Poskanzer and Sergei Voloshin
for their stimulating criticism.

%\newpage
\vspace{1cm}
\begin{center}
{\bf Figure Caption}
\end{center}
\vspace{0.5cm}

\noindent
{\bf Fig. 1.} 
$\Phi-$measure of $\phi-$fluctuations in the hadron gas as a function 
of temperature for four values of the chemical potential. The resonances are 
either neglected (dashed lines) or taken into account (solid lines). The most 
upper dashed and solid lines correspond to $\mu = 70$ MeV, the lower ones 
to $\mu = 0$, etc.


\newpage

\begin{thebibliography}{99}

\bibitem{Wan97} X.-N.~Wang, Phys. Rep. {\bf 280} (1997) 287.

\bibitem{Ame91} N.S.~Amelin et al., Phys. Rev. Lett. {\bf 67} (1991) 1523.

\bibitem{Oll92} J.-Y.~Ollitrault, Phys. Rev. {\bf D46} (1992) 229.

\bibitem{Rei97} W.~Reisdorf and H.G.~Ritter, Ann. Rev. Nucl. Part. Sci. 
{\bf 47} (1997) 663.

\bibitem{Agg97} M.M.~Aggarwal et al., Phys. Lett. {\bf B403} (1997) 390.

\bibitem{App98} H.~Appelsh\"auser et al., 
Phys. Rev. Lett. {\bf 80} (1998) 4136.

\bibitem{Mro93} St. Mr\' owczy\' nski, Phys. Lett. {\bf B314} (1993) 118;
{\it ibid.} {\bf B393} (1997) 26; Phys. Rev. {\bf C49} (1994) 2191.

\bibitem{Gyu97} M.~Gyulassy, D.~Rischke, and B.~Zhang, 
Nucl. Phys. {\bf A613}, 397 (1997).

\bibitem{Wan98} R.~Wang and H.~Sorge, Phys. Rev. {\bf C59} (1999) 1608.

\bibitem{Vol96} S.~Voloshin and Y.~Zhang, Z. Phys. {\bf  C70} (1996) 665.

\bibitem{Pos98} A.~Poskanzer and S.~Voloshin, Phys. Rev. {\bf  C58} (1998) 1671.

\bibitem{Oll97} J.-Y.~Ollitrault, nucl-ex/9711003.

\bibitem{Wan91} S.~Wang et al., Phys. Rev. {\bf C44} (1991) 1091.

\bibitem{Gaz92} M.~Ga\' zdzicki and St.~Mr\' owczy\' nski, 
Z. Phys. {\bf C54} (1992) 127.

\bibitem{Liu99} F.~Liu et al., Euro. Phys. J. {\bf C8} (1999) 649.

\bibitem{Rol98} G.~Roland and NA49 Collaboration, Nucl. Phys. 
{\bf A638} (1998) 91c.

\bibitem{App99} H.~Appelsh\"auser et al., Phys. Lett. {\bf B459} (1999) 679.

\bibitem{Mro98} St.~Mr\'owczy\'nski, Phys. Lett. {\bf B439} (1998) 6.

\bibitem{Mro99a} St.~Mr\'owczy\'nski, Phys. Lett. {\bf B459} (1999) 13.

\bibitem{Mro99b} St.~Mr\'owczy\'nski, nucl-th/9905021,
to appear in Phys. Lett. {\bf B}.

\bibitem{Sol91} J.~Sollfrank, P.~Koch and U.~Heinz, 
Z. Phys. {\bf C52} (1991) 593.

\bibitem{App98a} H.~Appelsh\"auser et al., 
Euro. Phys. J. {\bf C2} (1998) 661.

\bibitem{Bel99} M.~Belkacem et al., nucl-th/9903017.

\end{thebibliography}
\end{document}